\documentclass[prc,twocolumn,showpacs]{revtex4}
\usepackage{amssymb}
\usepackage{epsfig}     %{epsf}
\usepackage{wrapfig}
\usepackage{graphics}
\usepackage{amssymb}
\begin{document}

\title{Dipole polarizabilities of charged pions }

\author{L.V.~Fil'kov$^1$\footnote[1]{filkov@sci.lebedev.ru} and
V.L.~Kashevarov$^{2,1}$  
\vspace*{0.3cm}}

\affiliation{$^1$Lebedev Physical Institute, 119991 Moscow, Russia\\
$^2$Institut f\"ur Kernphysik, Johannes Gutenberg-Universit\"at Mainz,
55099 Mainz, Germany}

%======================================================================
\begin{abstract}
We discuss main experimental works, where dipole polarizabilities of
charged pions have been determined. 
Possible reasons for the differences between the experimental data are 
discussed. In particular, it is shown that the account of the
$\sigma$-meson gives a significant correction to the value of the 
polarizability obtained in the latest experiment of 
the COMPASS collaboration.
\end{abstract}

\pacs{13.40.-f, 11.55.Fv, 11.55.Hx, 12.39.Fe, 14.40.-n}

\maketitle                                                                               

%======================================================================
\section{Introduction}

Pion polarizabilities are fundamental structure parameters characterizing
the behavior of the pion in an external electromagnetic field.
The dipole ($\alpha_1$ and $\beta_1$) and quadrupole ($\alpha_2$ and $\beta_2$) 
polarizabilities are defined \cite{rad,fil2} through the expansion of the 
non-Born helicity amplitudes of the Compton scattering on the pion over $t$ at 
the fixed $s=\mu^2$:
%---------------
\begin{eqnarray}
\label{mpm}
&& \qquad \qquad  M_{++}(s=\mu^2,t)\nonumber \\
&&=\pi\mu\left[2(\alpha_1-\beta_1)+\frac{t}{6}(\alpha_2-\beta_2)\right]
+{\cal O}(t^2), 
\end{eqnarray}
\begin{eqnarray}
&& \qquad \qquad M_{+-}(s=\mu^2,t)\nonumber \\
&&=\frac{\pi}{\mu} \left[2(\alpha_1+\beta_1)+\frac{t}{6}(\alpha_2+\beta_2)
\right]+{\cal O}(t^2),\nonumber  
\end{eqnarray} 
%---------------
where $s$ ($t$) is the square of the total energy (momentum transfer) in the
$\gamma\pi$ center of mass (c.m.) system, $\mu$ is the pion mass, $\alpha_i$ 
and $\beta_i$ are electric and magnetic polarizabilities correspondingly.
In the following the dipole polarizabilities are given in units 
$10^{-4}\,{\rm fm}^3$. 

The values of the pion polarizabilities are very sensitive to predictions of
different theoretical models. 
Therefore, an accurate experimental determination of these parameters is very 
important for testing the validity of such models.

The most of experimental data obtained for the difference of the dipole 
polarizabilities of the charged pions are presented in Table I. 

The polarizabilities were determined by analyzing the processes of the high 
energy pions scattering in the Coulomb field of heavy nuclei 
$(\pi^-A\to\gamma\pi^-A')$ via the Primakoff effect,
radiative pion photoproduction from proton $(\gamma p\to\gamma\pi^+n)$, and 
two-photon production of pion pairs $(\gamma\gamma\to\pi\pi)$.
As seen from Table I, the data vary from 4 up to 40 and are in conflict even 
for experiments performed with the same method.
In this paper we will consider possible reasons for such disagreements.

%**************************** Table I *****************************************
\begin{table*}
\caption{Review of experimental data on $(\alpha_1-\beta_1)_{\pi^{\pm}}$} 
\centering
\begin{tabular}{|ll|l|} \hline
\multicolumn{2}{|c|}{Experiments} & \qquad\quad $(\alpha_1-\beta_1)_{\pi^{\pm}}$ 
\\ \hline
$\gamma p\to\gamma\pi^+n$ & MAMI (2005) \cite{mami} 
& $11.6\pm 1.5_{stat}\pm 3.0_{syst}\pm 0.5_{mod}$ \\ \hline
$\gamma p\to\gamma\pi^+n$ & Lebedev Phys. Inst. (1984) \cite{lebed}& 
$40\pm 20$ \\ \hline
$\pi^-A\to\gamma\pi^-A'$ & Serpukhov (1983) \cite{serp} & $13.6\pm 2.8 \pm2.4$ 
\\ \hline  
$\pi^-A\to\gamma\pi^-A'$ & COMPASS (2007) \cite{comp} & $4.0\pm 1.2\pm 1.4 $ 
\\ \hline

\multicolumn{2}{|c|}
{D. Babusci {\em et al.} (1992) \cite{bab}} &   \\
     & PLUTO \cite{pluto}   & $38.2\pm 9.6\pm 11.4$ \\
     & DM 1 \cite{dm1}     & $34.4\pm 9.2$  \\
     & MARK II \cite{mark} & $4.4\pm 3.2$    \\ \hline
\multicolumn{2}{|c|}
{J.F. Donoghue, B.R. Holstein (1993) \cite{don}} &  5.4 \\
\multicolumn{2}{|c|}  
{MARK II \cite{mark}}  &       \\ \hline
\multicolumn{2}{|c|}  
{A.E. Kaloshin, V.V. Serebryakov (1994) \cite{kal}} & $5.25\pm 0.95$ \\
\multicolumn{2}{|c|}
{MARK II \cite{mark}}   &         \\ \hline\hline
\multicolumn{2}{|c|}
{L.V. Fil'kov, V.L. Kashevarov (2006) \cite{fil3}}&  $13^{+2.6}_{-1.9}$ \\
$\gamma\gamma\to\pi^+\pi^-$ & fit of data \cite{mark,tpc,cello,ven,aleph,belle}& 
 \\
       & from threshold to 2.5 GeV  &  \\ \hline
\multicolumn{2}{|c|}
{R. Garcia-Martin, B. Moussallam} &  \\
 & (2010) \cite{garcia}, $\gamma\gamma\to\pi^+\pi^-$  & 4.7 \\  \hline
 
\end{tabular}
\end{table*}
                                                                                      
%======================================================================
\section{Radiative photoproduction of the $\pi^+$-meson from the proton}

An experiment on the radiative photoproduction $\pi^+$-meson from the proton 
$(\gamma p\to\gamma\pi^+n)$ was carried out at the Mainz Microtron MAMI 
\cite{mami} in the kinematical 
region of 540 MeV $< E_{\gamma} <$ 820 MeV and 
$140^{\circ}\leq \theta^{cm}_{\gamma\gamma}\leq 180^{\circ}$, where 
$\theta^{cm}_{\gamma\gamma}$ is a polar angle in the c.m. system of the outgoing 
photon and pion. 

The theoretical calculations of the cross section for the reaction
$\gamma p\to\gamma\pi^+n$ show that the contribution of
nucleon resonances is suppressed for photons
scattered backward in the c.m. system of the reaction $\gamma\pi\to\gamma\pi$.
Moreover, integration over $\varphi$ and $\theta_{\gamma\gamma}^{cm}$ 
essentially decreases the contribution of nucleon resonances from the crossed 
channels. In addition, the difference $(\alpha_1-\beta_1)_{\pi^{\pm}}$ gives 
the biggest contribution to the cross section for $\theta_{\gamma\gamma}^{cm}$ 
in the same region of $140^{\circ}-180^{\circ}$. Therefore, one considered 
the cross section of radiative pion photoproduction integrated over $\varphi$ 
from $0^{\circ}$ to $360^{\circ}$ and over $\theta_{\gamma\gamma}^{cm}$ 
from $140^{\circ}$ to $180^{\circ}$,
%---------------
\begin{equation}
\int_0^{360^{\circ}}d\varphi\int_{-1}^{-0.766}d\cos\theta_{\gamma\gamma}^{cm}\;
\frac{d\sigma_{\gamma p\to\gamma\pi^+n}}{dt ds d\Omega_{\gamma\gamma}}.
\end{equation}
%---------------

The work was carried out at values of $s$ up to 15$\mu^2$. 
It has been shown in Ref. \cite{fil1}, that the contribution of the 
$\sigma$-meson is noticeable at such high values of $s$. The contributions 
of other mesonic resonances $(\rho, a_1, b_1, a_2)$ are negligible here.                                            

The values of the pion polarizabilities have been obtained from a fit
of the cross section calculated 
by two different theoretical models to the data.
%The cross section of the process $\gamma p\to\gamma\pi^+n$ has been
%calculated 
%in the framework of two different models.
In the first model the contribution of all
the pion and nucleon pole diagrams was taken into account.
In the second model in addition to the nucleon and the pion pole diagrams
(without the anomalous magnetic moments of nucleons) the contribution 
of the resonances $\Delta (1232)$, 
$P_{11}(1440)$,$D_{13}(1520)$, and $S_{11}(1535)$ and the $\sigma$-meson
were included. 

It should be noted that the contribution of the sum of the pion
polarizabilities is very small in the considered region of
$140^{\circ} \lesssim \theta_{\gamma\gamma}^{cm} \lesssim 180^{\circ}$ .
The estimate shows that the contribution of 
$(\alpha_1+\beta_1)_{\pi^{\pm}}=0.4$ to
the value of $(\alpha_1-\beta_1)_{\pi^{\pm}}$ is less than 1\%.

To increase the confidence that
the model dependence of the result was under control, 
kinematic regions were considered  where the difference between
the models did not exceed $3\%$ when $(\alpha_1-\beta_1)_{\pi^{\pm}}$ is 
constrained to zero.
First, the kinematic region was considered where the contribution of the
pion polarizability is negligible, i.e. the region
$1.5\mu^2 < s<5\mu^2$.
Then the kinematic region was investigated where the polarizability
contribution is biggest. This is the region $5\mu^2< s<15\mu^2$ and
$-12\mu^2<t<-2\mu^2$.
In the range $t>-2\mu^2$ the polarizability contribution is small
and this region was excluded.

Analysis of these data gave the following result
\begin{equation}\label{ab}
(\alpha_1-\beta_1)_{\pi^{\pm}}= 11.6\pm 1.5_{stat}\pm 3.0_{syst}\pm 0.5_{model}.     
\end{equation}

An independent analysis \cite{igor} of the experimental data
was carried out by a constrained $\chi^2$ fit. The result \cite{igor}
agree very well with (\ref {ab}) giving it additional support.

The result \cite{mami} is consisted with earlier works investigating the
$\gamma p\to\gamma\pi^+n$ \cite{lebed} and $\pi^-A\to\gamma\pi^-A'$ \cite{serp} 
reactions, and also with \cite{fil3}, where a global fit to all existing data 
for the $\gamma\gamma\to\pi^+\pi^-$ reaction was done.
On the other hand,
the result \cite{mami} is in conflict with the prediction of ChPT 
\cite{gasser2,burgi}. This discrepancy can be connected with a different
account of the contribution of the $\sigma$-meson and vector mesons in the
dispersion relations and ChPT calculations \cite{bern}.

%======================================================================
\section{Scattering of pions in the Coulomb field of heavy nuclei}

The first experimental data on the charged pion polarizability was
obtained in the work \cite{serp}. They studied the scattering of high 
energy $\pi^-$ mesons off the Coulomb field of heavy nuclei. A connection of
the radiative scattering in the Coulomb field with the Compton scattering
was first predicted in the work \cite{pom}.
                                                              
The cross section of the radiative pion scattering $\pi A\to\pi\gamma A'$  
via the Primakoff effect can be written as
%---------------
\begin{eqnarray}
\label{cross}
&& \qquad \qquad \frac{d\sigma_{\pi A}}{ds dQ^2 d \cos\theta_{\gamma\gamma}^{cm}}
\\
&&=\frac{Z^2 \alpha}{\pi (s-\mu^2)} 
 F^2_{eff}(Q^2)\frac{Q^2-Q^2_{min}}{Q^4} \frac{d\sigma_{\pi\gamma}}
{d\cos\theta_{\gamma\gamma}^{cm}},
\nonumber 
\end{eqnarray} 
%--------------------
where $F_{eff}\approx 1$ is the electromagnetic form-factor of nucleus,
$\alpha$ is the fine-structure constant, $Z$ is the charge number of 
the nucleus,
and $Q^2$ is the negative 4-momenta transfer squared, $Q^2=-(p_A-p_A')^2$.
$Q^2_{min}$ is the minimum value of $Q^2$ which is given by the formula
%---------------
\begin{equation}                          
Q^2_{min}=\frac{(s-\mu^2)^2}{4 E^2_{beam}},
\label{min}
\end{equation}
%---------------
where $s$ is the square of the total energy of the process 
$\gamma+\pi^{\pm}\to\gamma+\pi^{\pm}$,
$E_{beam}$ is the pion beam energy.  

This cross section has a peak at $Q^2=2 Q^2_{min}$ with a width equal to
$\simeq 6.8 Q^2_{min}$.

The experiment \cite{serp} was carried at a beam energy equal to 40 GeV.
In this case if the energy of the incident photon in the incident pion rest 
frame $\omega_1=600$ MeV, then $Q^2_{min}$ is equal to $4.4\times 10^{-6}$
(GeV/c)$^2$. It was shown that the Coulomb amplitude dominates in this case 
for $Q^2 \leq 2\times 10^{-4}$(GeV/c)$^2$. The experiment \cite{serp} was 
carried out at $Q^2_{cut}<6\times 10^{-4}$(GeV/c)$^2$. Events in the region of 
$Q^2$ of $(2-8)\times 10^{-3}$(GeV/c)$^2$ were used for estimation of the 
strong interaction background. This background was assumed to behave either
as $\sim Q^2$ in the region $Q^2 \leq 6\times 10^{-4}$(GeV/c)$^2$ or as
a constant. The polarizability was determined from the ratio (assuming
$(\alpha+\beta)_{\pi^{\pm}}=0$)
%---------------
\begin{equation}
\label{R}
R_{\pi}=(\frac{d\sigma_{\gamma\pi}}{d\Omega})/(\frac{d\sigma_{\gamma\pi}^0}
{d\Omega})=1-\frac{3}{2}\frac{\mu^3}{\alpha}
\frac{x_{\gamma}^2}{1-x_{\gamma}}\alpha_{\pi},
\end{equation}
%---------------
where $\frac{d\sigma_{\gamma\pi}}{d\Omega}$ refers to the measured cross 
section and $\frac{d\sigma_{\gamma\pi}^0}{d\Omega}$ to simulated 
cross section expected for $\alpha_{\pi}=0$, $x_{\gamma}=E_{\gamma}/E_{beam}$ 
in the laboratory system of the process $\pi A\to\pi\gamma A'$. 
As a result they have obtained 
%---------------
\begin{equation}
 (\alpha_1-\beta_1)_{\pi^{\pm}}=13.6 \pm 2.8 \pm 2.4 .
\end{equation}
%---------------

The new result of the COMPASS collaboration \cite{comp} for the charged pion 
electric polarizability $\alpha_{\pi}=2.0\pm 0.6_{stat.}\pm 0.7_{syst.}$ has 
been found also by studying the $\pi^-$-meson scattering off the Coulomb field 
of heavy nuclei. The result was obtained assuming that $\alpha_1=-\beta_1$. 
This value is at variance with the result  obtained in a very similar 
experiment in Serpukhov \cite{serp}, but also with \cite{mami}. 
 
The COMPASS experiment \cite{comp} was performed with $E_{beam}=190$ GeV.  
%assuming that $\alpha_1=-\beta_1$. 
For such values of $E_{beam}$ the quantity of $Q^2_{min}(COMPASS)$ must be 
smaller by 22.5 times than $Q^2_{min}$(Serpukhov). In this experiment 
\cite{comp} the authors considered $Q^2_{cut}\lesssim 0.0015$ (GeV/c)$^2$. 

As shown in \cite{faldt1} the basic ratio $R_{\pi}$ is applicable for the 
Coulomb peak only. In Ref.~\cite{faldt2} it is elaborated that the Coulomb 
amplitude interference with the coherent nuclear amplitude is important for 
$0.0005 \leqslant Q\leqslant 0.0015$\,(GeV/c)$^2$. 
This means that the Serpukhov analysis could safely apply the ratio $R_{\pi}$ 
in (\ref {R}), whereas COMPASS has to consider the interference of the Coulomb 
and strong amplitude. The phase determined with the simple considerations in 
Ref.~\cite{walch} for the Serpukhov experiment \cite{serp} is close to $\pi/2$ 
meaning that the subtraction of a nuclear background assumed to be incoherent 
is justified. In the COMPASS analysis a Gaussian profile function is used for 
the diffractive background 
%\cite{comprivcom} 
and the relative phase is 
determined by a fit to the cross sections. The contributions to the fit are 
not shown in Fig.~3(c) of Ref.~\cite{comp}. With a more realistic 
"absorbing disc" for the profile function \cite{amaldi} all 
bumps in Fig.~3(c) are well reproduced and again a phase is close to $\pi/2$
is indicated \cite{WalchPrivat}.
Without a real fit to the data it is impossible to estimate the 
effect of the model dependence of the diffractive background, but that it will 
have an influence is clear from Ref.~\,\cite{faldt2}. Considering this situation 
the discussion of the central subtraction of the diffractive background is 
insufficient to get a feeling for model dependence of the analysis.

Comparison of data with different targets provide the possibility to check
the $Z^2$ dependence for the Primakoff cross section and to estimate a possible
contribution of the nuclear background. Such an investigation was performed
by the Serpukhov collaboration and they have obtained $Z^2$ dependence with
good enough accuracy. The COMPASS collaboration really have gotten their main
result using only $Ni$ target but they wrote that they also considered other 
targets on small statistic and obtained approximate $\sim Z^2$  dependence.

It should be noted that in order to get an information about the pion 
polarizabilities, the authors considered the cross section of the process 
$\gamma\pi^-\to\gamma\pi^-$ equal to the Born cross section and the interference 
of the Born amplitude with the pion polarizabilities only. The COMPASS 
collaboration analyzed this process up to the total energy $W=490$ MeV in the 
angular range $0.15>\cos\theta^{cm}_{\gamma\pi}>-1$. However, the contribution 
of the $\sigma$-meson to the cross section of the Compton scattering on the pion 
could be very substantial in this region of the energy and angles.  Therefore, 
we consider this contribution.

%======================================================================
\section{$\sigma$-meson contribution}

The cross section of the elastic $\gamma\pi$ scattering can be written as 
\cite{fil1}:
\begin{eqnarray}
\label{siggp}
\lefteqn{\frac{d\sigma_{\gamma\pi\to\gamma\pi}}{d\Omega} = 
\frac{1}{256\pi^2}\frac{(s-\mu^2)^4}{s^3}} \nonumber  \\ 
 &&\times\left[(1-z)^2|M_{++}|^2+s^2(1+z)^2|M_{+-}|^2\right],
\end{eqnarray}

%-------------------
where $z=\cos\theta^{cm}_{\gamma\pi}$.
The amplitudes $M_{++}$ and $M_{+-}$ have no kinematical singularities and zeros
\cite{gold}. 

The dispersion relation (DR) for the amplitude $M_{++}$ 
at fixed $t$ with one subtraction was obtained in \cite{fil1}:
%---------------
\begin{eqnarray}
\label{ds}
\lefteqn{Re M_{++}(s,t)=Re \overline{M}_{++}(s=\mu^2,t)+B_{++}} \nonumber \\
 &&+\frac{(s-\mu^2)}{\pi}P~\int\limits_{4\mu^2}^{\infty}d s'~Im M_{++}(s',t)\\
&& \times\left[\frac{1}{(s'-s)(s'-\mu^2)}-\frac{1}{(s'-u)(s'-\mu^2+t)}\right],
\nonumber
\end{eqnarray}
%---------------
where $B_{++}$ is the Born term equal to
%---------------
\begin{equation}
B_{++}=\frac{2e^2\mu^2}{(s-\mu^2)(u-\mu^2)}.
\end{equation}
%---------------

Via the cross symmetry this DR is identical to a DR with two subtraction.
The subtraction function $Re \overline{M}_{++}(s=\mu^2,t)$
was determined with help of the DR at fixed $s=\mu^2$ with one
subtraction where the subtraction constant was expressed
through the difference $(\alpha_1-\beta_1)_{\pi^{\pm}}$:
%---------------
\begin{eqnarray}
\label{sub}
\lefteqn{Re\overline{M}_{++}(s=\mu^2,t)=2\pi\mu
(\alpha_1-\beta_1)_{\pi^{\pm}}} \nonumber \\ 
&&+\frac{t}{\pi}\left\{P\int\limits_{4\mu^2}^{\infty}
 \frac{Im M{++}(t',s=\mu^2)~d t'}{t'(t'-t)}\right. \\
&& -\left. P\int\limits_{4\mu^2}^{\infty}
 \frac{Im M_{++}(s',u=\mu^2)~d s'}{(s'-\mu^2)(s'-\mu^2+t)}\right\}\nonumber.
\end{eqnarray}
%---------------

The DRs for the amplitude $M_{+-}(s,t)$ have the same expressions (\ref{ds}) 
and (\ref{sub}) with substitutions: 
$Im M_{++} \to Im M_{+-}$, $B_{++} \to B_{+-}=B_{++}/\mu^2$ and
$2\pi\mu(\alpha_1-\beta_1)_{\pi^{\pm}} \to (2\pi/\mu) 
(\alpha_1+\beta_1)_{\pi^{\pm}}$.
 
The amplitudes $M_{++}$ and $M_{+-}$ were calculated with help of these DRs 
taking into account contribution of the following mesons: $\rho (770)$, 
$b_1 (1235)$, $a_1 (1260)$, and $a_2 (1320)$ mesons  in the $s$-channel and 
$\sigma(600)$, $f_0 (980)$, $f'_0 (1370)$, $f_2 (1270)$ mesons in the 
$t$-channel. It has been shown that the contribution of all these mesons, 
with exception of the $\sigma$-meson, is very small in the region of the energy 
and the angles of the COMPASS experiment.

According DRs (\ref{sub}) the contribution of the $\sigma$-meson can be 
determined as
%---------------
\begin{equation}
Re M_{++}^{\sigma} = \frac{t}{\pi}\int\limits_{4\mu^2}^{\infty}~\frac{
Im M_{++}^{\sigma}(t',s=\mu^2)~d t'}{t'(t'-t)}.
\end{equation}
%---------------

The imaginary amplitude $Im M_{++}^{\sigma}(t,s=\mu^2)$ has to be evaluated 
taking into account that the $\sigma$-meson is a pole on the second Riemann 
sheet. The relation between amplitudes on the first and the second sheets can 
be written \cite{oller} as
%---------------
\begin{equation}
\label{f0}
F_0^{II}(t+i\epsilon)=F_0^{I}(t+i\epsilon)(1+2i\rho T_0^{II}(t+i\epsilon)),
\end{equation} 
%---------------
where
%---------------
\begin{equation}                                                          
T_0^{II}=-\frac{g_{\sigma\pi\pi}^2}{t_{\sigma}-t}, \quad  
F_0^{II}=\sqrt{2}\,\frac{g_{\sigma\gamma\gamma} g_{\sigma\pi\pi}}{t_{\sigma}-t},
\end{equation}
%---------------
%---------------
\begin{eqnarray}
&& t_{\sigma}=(M_{\sigma}-i\Gamma_{\sigma 0}/2)^2, \quad  
\rho=\frac{\sqrt{1-4\mu^2/t}}{16\pi},\\
&&\qquad \Gamma_{\sigma 0}=\Gamma_{\sigma}\left( \frac{t-4\mu^2}
{M_{\sigma}^2-4\mu^2}\right)^{1/2}.\nonumber
\end{eqnarray}
%---------------

Using the relation (\ref{f0}) we have
%---------------
\begin{equation}
\label{im}
Im M_{++}^{\sigma}(t,s=\mu^2)=\frac{1}{t}\sqrt{\frac{2}{3}}\,
\frac{g_{\sigma\gamma\gamma} g_{\sigma\pi\pi}R}{D^2+R^2},
\end{equation}
%---------------
where
%---------------
\begin{equation}
D=(M_{\sigma}^2-t-\frac{1}{4}\Gamma_{\sigma 0}^2), \quad 
R=M_{\sigma}\Gamma_{\sigma 0}+2\rho g_{\sigma\pi\pi}^2.   
\end{equation}
%---------------

We can get influence of the $\sigma$-meson on the extracted value of 
$(\alpha_1-\beta_1)_{\pi^{\pm}}$
by equating the cross section  without $\sigma$-meson contribution to the
cross section when $\sigma$-meson is taking into account:
%---------------
\begin{eqnarray}
\label{rel}
&& \qquad d\sigma_{\gamma\pi\to\gamma\pi}(B, (\alpha_1-\beta_1)_{\pi^{\pm}}^0)/
d\Omega \nonumber \\
&& =\sigma_{\gamma\pi\to\gamma\pi}(B, M_{++}^{\sigma} 
(\alpha_1-\beta_1)_{\pi^{\pm}})/d\Omega,
\end{eqnarray}
%---------------
where
$(\alpha_1-\beta_1)_{\pi^{\pm}}^0$ is the value of 
$(\alpha_1-\beta_1)_{\pi^{\pm}}$ without of the 
$\sigma$ contribution obtained in \cite{comp}) and $B$ is the Born term.

For backward scattering ($z=-1$), we have the following expression:
%---------------
\begin{eqnarray}
\label{amc1}
&& \qquad \qquad(\alpha_1-\beta_1)_{\pi^{\pm}} \\ 
&& =\frac{1}{4\pi\mu}\left\{-(B+Re M_{++}^{\sigma}) 
     +\frac{B^2+4\pi\mu B(\alpha_1-\beta_1)_{\pi^{\pm}}^0}
{B+Re M_{++}^{\sigma}}\right\}.
\nonumber
\end{eqnarray}  
%---------------

%**************************** Fig. 1 *****************************************
\begin{figure}
\epsfxsize=6cm      
\epsfysize=7cm      
\centerline{              
\epsffile{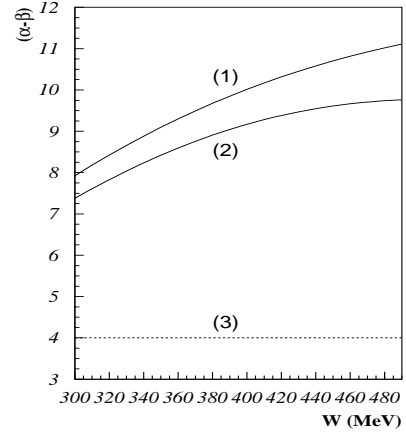}}              
\caption{Dependence of $(\alpha_1-\beta_1)_{\pi^{\pm}}$ on $W$. Lines (1) 
and (2) correspond to the calculation Eqs. (\ref{amc1}) and  (\ref{amc2}), 
respectively. Line (3) is the COMPASS result \cite{comp}.}
\label{(a-b)1}
\end{figure}
%*******************************************************************************

In the case of integration over the region $-1\leq z\leq 0.15$ we have
%---------------
\begin{equation}
\label{amc2}
(\alpha_1-\beta_1)_{\pi^{\pm}}=F_0/F_1,           
\end{equation}
%---------------
where
%--------------
\begin{eqnarray}
\label{f01}            
&&F_0 =\frac{1}{4\pi\mu}\left\{\int\limits_{-1}^{0.15}~\left[-Re M_{++}^{\sigma}
  (Re M_{++}^{\sigma}+2 B)\right.\right. \nonumber \\ 
&&+\left.\left. 4\pi\mu B(\alpha_1-\beta_1)_{\pi^{\pm}}^0\right](1-z)^2~dz 
\right\},
\end{eqnarray} 
%--------------
\begin{equation}
\label{f1}
F_1=\left\{\int\limits_{-1}^{0.15}~(B+Re M_{++}^{\sigma})(1-z)^2~dz\right\}.
\end{equation}     
%--------------
In the calculation we used the parameters of the $\sigma$-meson from 
Ref.~\cite{oller}:
$M_{\sigma}=441$MeV, $\Gamma_{\sigma}=544$MeV, 
$\Gamma_{\sigma\gamma\gamma}=1.98$keV, $g_{\sigma\pi\pi}=3.31$GeV, 
$g_{\sigma\gamma\gamma}^2=16\pi\Gamma_{\sigma\gamma\gamma}$.       
The results of the calculations using Eq.~(\ref{amc1}) (line (1) ) 
and Eq.~(\ref{amc2}) (line (2)) are shown in Fig.~1. Line (3) is the result of 
Ref.~\cite{comp}. As a result we obtain 
$(\alpha_1-\beta_1)_{\pi^{\pm}} \sim 10$. However the
magnitude of $(\alpha_1-\beta_1)_{\pi^{\pm}}$ is very sensitive to parameters 
of the $\sigma$-meson and can reach a value of $\sim 11$ for the parameters from 
\cite{penn}.

So, the contribution of the $\sigma$-meson can essentially change the COMPASS 
result. It should be noted that the contribution of the 
$\sigma$-meson was not considered in Serpukhov as well. 
However, in this case, the contribution of the $\sigma$-meson for the Serpukhov
kinematics is $\Delta (\alpha-\beta)_{\sigma}\gtrsim 2.7$ within the
experimental error of the Serpukhov result.

%======================================================================
\section{Two-photon production of pion pairs}
                                      
The information about pion polarizabilities could be obtained also  by
studying the cross section of the reaction $\gamma\gamma\to\pi^+\pi^-$.

%**************************** Fig. 2 *******************************************
\begin{figure*}[t]\label{crosy}
\epsfxsize=11.5cm
\epsfysize=10.0cm
\centerline{
\epsffile{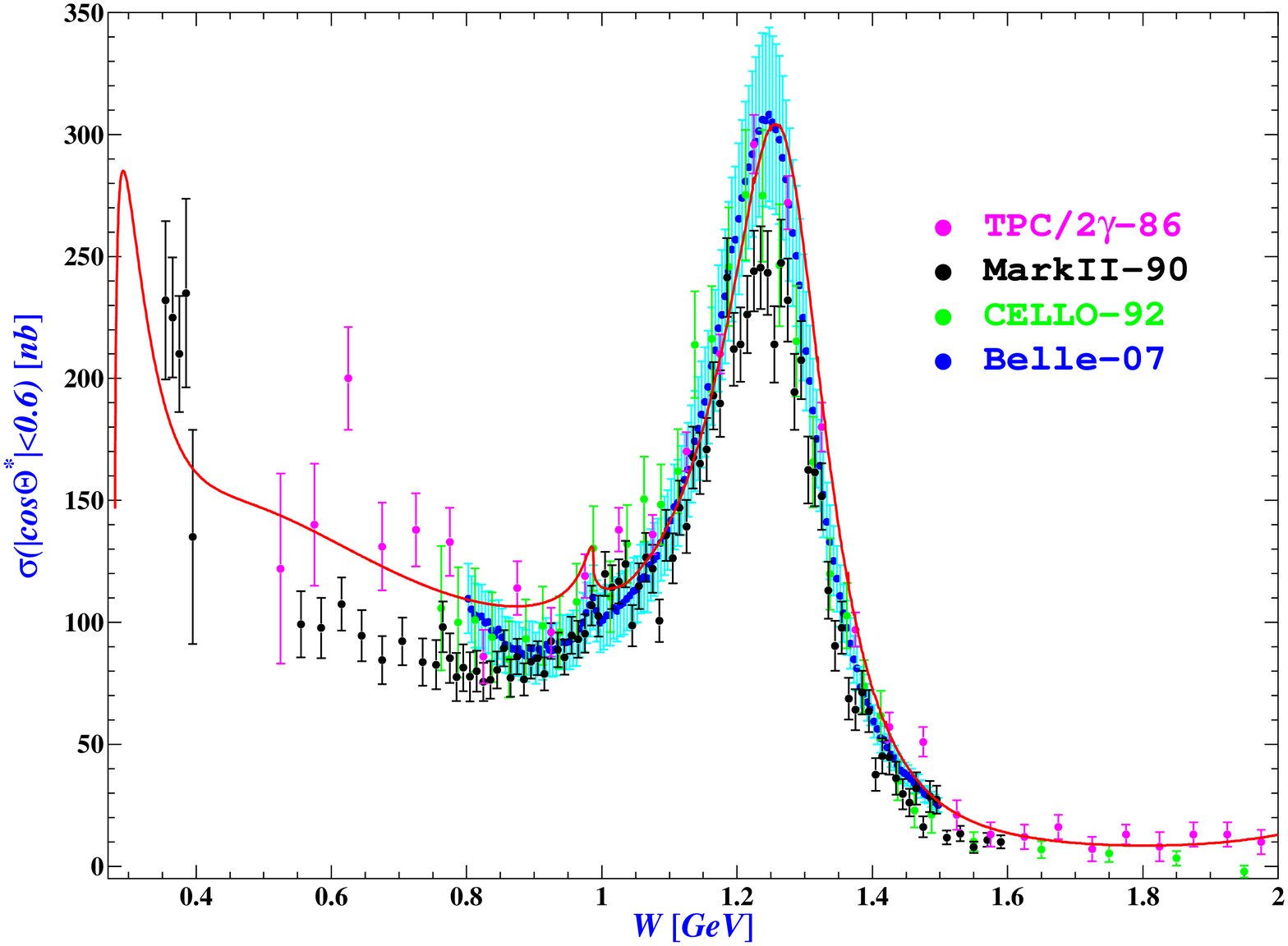}}
\caption{
The cross section of the process $\gamma\gamma\to\pi^+\pi^-$ 
(with $\cos\theta_{cm}<0.6)$. Experimental data of TPC/2$\gamma$ \cite{tpc}, Mark II \cite{mark},
and CELLO \cite{cello} Collaborations are shown with statistical uncertainties only. 
Statistical uncertainties for the most of the Belle Collaboration data \cite{belle} are smaller 
then corresponding blue circles. Vertical light-blue error bars are systematic uncertainties 
for these data.    
}
\end{figure*}
%*******************************************************************************

Investigation of this process at low and middle energies was carried
out in the frameworks of different theoretical models
%\cite{kal,bab,bij,don,ser,bell,oset1,fil1,lee,penn1,drechs,penn2,oller2,
%mao,garcia,hofer}
and, in particular, within dispersion relations.

Authors of most dispersion approaches used DRs for partial
waves taking into account the contribution of $S$ and $D$ wave
only. Moreover, they often used
additional assumptions, for example, to determine  subtraction
constants. The dipole polarizabilities of charged pions were obtained in
works \cite{kal,don,bab,bij,garcia,hofer} from the analysis of the experimental
data in the region of the low energy ($ W<700$ MeV) mainly (where $W$ is the
total energy in $\gamma\gamma$ c.m. system). 
%The values of the polarizabilities
The most of results for the charged pion polarizabilities
obtained in these works are close to the ChPT prediction \cite{gasser2,burgi}.
On the other hand, the values
of the experimental cross section of the process $\gamma\gamma\to\pi^+\pi^-$ 
in this region
are very ambiguous. Therefore, as it
has been shown in Ref.~\cite{don,fil3},
even changes of these values by more than 100\% are still
compatible with the present error bars in the energy region considered.
Therefore at present, more realistic values of the polarizabilities could 
be obtained analyzing the experimental data on $\gamma\gamma\to\pi^+\pi^-$ in a 
wider energy region.

The processes $\gamma\gamma\to\pi^0\pi^0$ and $\gamma\gamma\to\pi^+\pi^-$ 
were analyzed in Ref.~\cite{fil2,fil3,fil1}
using DRs with subtraction for the invariant amplitudes $M_{++}$ and
$M_{+-}$ without an expansion over partial waves. The subtraction constants
are uniquely determined in these works through the pion polarizabilities.
The values of polarizabilities have been found from the fit to the
experimental data of the processes $\gamma\gamma\to\pi^+\pi^-$ and 
$\gamma\gamma\to\pi^0\pi^0$ up to 2500 MeV and
2250 MeV, correspondently. As a result the following  values of
$(\alpha_1-\beta_1)_{\pi^{\pm}}=13.0^{+2.6}_{-1.9}$ and 
$(\alpha_1-\beta_1)_{\pi^0}=-1.6\pm 2.2$ have been
found in these works. In addition, for the first time there were obtained 
quadrupole polarizabilities for both charged and neutral pions.   

The new result of the calculation of the total cross section of the process
$\gamma\gamma\to\pi^+\pi^-$ at $\cos\theta^{cm}_{\gamma\gamma}<0.6$ in the 
frame of the DRs  
\cite{fil3} has been obtained with
the $\sigma$-meson considered as a pole on the second Riemann sheet.
The result,
using Eq.~(\ref{im}) for $Im M_{++}^{\sigma}(t,s=\mu^2)$ with
the following parameters of the $\sigma$-meson: $M_{\sigma}=441$ MeV,
$\Gamma_{\sigma}=544$ MeV, 
$\Gamma_{\sigma\gamma\gamma}=1 $keV, $g_{\sigma\pi\pi}=2.924$GeV
$(\alpha_1-\beta_1)_{\pi^{\pm}}=10$,
is shown in Fig.~2.

However, the region $W\lesssim 800$ MeV is most sensitive to the contribution
of $(\alpha_1-\beta_1)_{\pi^{\pm}}$. Therefore,
in order to obtain real values of the polarizabilities of the
charged pions, it is necessary to have new more accurate data for the process
$\gamma\gamma\to\pi^+\pi^-$ in the energy region $W<800$ MeV. It should be 
noted that even in the
region of $W\lesssim 500$ MeV the main contribution is given by the Born term,
the dipole and quadrupole polarizabilities, and the $\sigma$-meson. 

%-----------------------------------------------------------------------------
In conclusion, further experimental and theoretical investigations are
needed to determine the true value of the pion polarizabilities.
The authors thanks Th. Walcher and A.I. L'vov for useful discussions.

%%%%%%%%%%%%%%%%%%%%%%%%%%%%%%%%%%%%%%%%%%%%%%%%%%%%%%%%%%%%%%%%%%%%%%%%%%%%%%%%%%%%%

%%%%%%%%%%%%%%%%%%%%%%%%%%%%%%%%%%%%%%%%%%%%%%%%%%%%%%%%%%%%%%%%%%%%%%%%%%%%%%%%%%%%%

\end{document}